\documentclass[prl,twocolumn,groupedaddress,showpacs,nofootinbib]{revtex4}
\usepackage{graphicx}
\usepackage{dcolumn}% Align table columns on decimal point
\usepackage{bm}% bold math
\usepackage{amssymb}
\usepackage{mathrsfs}
\usepackage{amsmath}
\usepackage{epsfig}
\usepackage[dvips]{color}
\usepackage{hhline}

\begin{document}

\title{Relations between Neutrino and Charged Fermion Masses}

\author{Pei-Hong Gu}
\email{peihong.gu@mpi-hd.mpg.de}

\affiliation{Max-Planck-Institut f\"{u}r Kernphysik, Saupfercheckweg
1, 69117 Heidelberg, Germany}

\begin{abstract}

We find an intriguing relation between neutrino and charged fermion
masses, $|m_{\nu_3^{}}^2- m_{\nu_1^{}}^2| : (m_{\nu_2^{}}^2-
m_{\nu_1^{}}^2):: V_{tb}^4 m_\tau^2 m_b^2/m_t^2 : V_{cs}^4 m_\mu^2
m_s^2/m_c^2$. We further indicate this relation can be predicted by
a left-right symmetric model.

\end{abstract}


\maketitle

The neutrino oscillation data have determined the values of the two
neutrino mass squared difference \cite{stv2008},
\begin{eqnarray}
~\Delta m_{21}^2~ &=&
~m_{\nu_2^{}}^2-m_{\nu_1^{}}^2~=7.65^{+0.23}_{-0.20} \times
10^{-5}_{}\,\textrm{eV}^2_{}\,,\nonumber\\
\vspace{10mm} |\Delta m_{31}^2| &=&
|m_{\nu_3^{}}^2-m_{\nu_1^{}}^2|\,=2.40^{+0.12}_{-0.11}\times
10^{-3}_{}\,\textrm{eV}^2_{}\,.
\end{eqnarray}
We notice the ratio of the above neutrino mass squared difference
can be well described by the quark and charged lepton masses,
\begin{eqnarray}
\label{relation} |\Delta m_{31}^2| : \Delta m_{21}^2 ::
\frac{m_b^2}{m_t^2}m_\tau^2 V_{tb}^4 : \frac{m_s^2}{m_c^2}m_\mu^2
V_{cs}^4\,,
\end{eqnarray}
with the charged fermion masses \cite{xzz2007},
\begin{eqnarray}
\label{fermionmass}
m_u^{}&=&1.27^{+0.50}_{-0.42}\,\textrm{MeV}\,,~~~~~~~m_d^{}=2.90^{+1.24}_{-1.19}\,\textrm{MeV}\,,\nonumber\\
m_c^{}&=&0.619\pm0.084\,\textrm{GeV}\,,~~m_s^{}=55^{+16}_{-15}\,\textrm{MeV}\,,\nonumber\\
m_t^{}&=&171.7\pm3.0\,\textrm{GeV}\,,~~~~~m_b^{}=2.89\pm0.09\,\textrm{GeV}\,,\nonumber\\
m_e^{}&=&0.486570161\pm0.000000042\,\textrm{MeV}\,,\nonumber\\
m_\mu^{}&=&102.7181359\pm0.0000092\,\textrm{MeV}\,,\nonumber\\
m_\tau^{}&=&1776.24^{+0.20}_{-0.19}\,\textrm{MeV}\,,
\end{eqnarray}
and the Cabibbo-Kobayashi-Maskawa (CKM) \cite{cabibbo1963} matrix
\cite{pdg2008},
\begin{eqnarray}
\label{ckm} &&V_{\textrm{CKM}}^{} =\left(\begin{array}{ccc}
V_{ud}^{}&
V_{us}^{}&V_{ub}^{}\\
[-2mm]~&~&~\\ V_{cd}^{}&
V_{cs}^{}&V_{cb}^{}\\
[-2mm]~&~&~\\ V_{td}^{}& V_{ts}^{}&V_{tb}^{}
\end{array}\right)=\nonumber\\
&&\hskip -1.5cm \left(\begin{array}{ccc}
0.97419\pm0.00022&0.2257\pm0.0010
&0.00359\pm0.00016\\
[-2mm]~&~&~\\ 0.2256\pm0.0010&
0.97334\pm0.00023 &0.0415^{+0.0010}_{-0.0011}\\
[-2mm]~&~&~\\ 0.00874^{+0.00026}_{-0.00037}&
0.0407\pm0,0010&0.999133^{+0.000044}_{-0.000043}
\end{array}\right)\,.\phantom{xxx}\nonumber
\end{eqnarray}
\vspace{-5mm}
\begin{eqnarray}
\end{eqnarray}
Here the charged fermion masses and the CKM matrix are all given at
$\mu=m_Z^{}$.

A possible solution to the intriguing relation (\ref{relation}) is
to consider the normal hierarchy neutrino masses,
\begin{eqnarray}
\label{spectrum} \hskip
-1cm\displaystyle{(m_{\nu_1^{}}^{}\,,~m_{\nu_2^{}}^{}\,,~m_{\nu_3^{}}^{})}
=\epsilon
\displaystyle{\left(\frac{m_d^{}}{m_u^{}}m_e^{}V_{ud}^2\,,~\frac{m_s^{}}{m_c^{}}m_\mu^{}V_{cs}^2\,,
~\frac{m_b^{}}{m_t^{}}m_\tau^{}V_{tb}^2\right)}\,,
\end{eqnarray}
where the parameter $\epsilon$ is defined by
\begin{eqnarray}
\label{parameter1}\epsilon&=&\sqrt{\frac{|\Delta m_{31}^2|}{
V_{tb}^4 m_\tau^2 m_b^2 /m_t^2-V_{ud}^4 m_e^2 m_d^2/m_u^2
}}\nonumber\\
&=&\sqrt{\frac{\Delta m_{21}^2} { V_{cs}^4 m_\mu^2
m_s^2/m_c^2-V_{ud}^4 m_e^2 m_d^2/m_u^2}}\,.
\end{eqnarray}
We take
\begin{eqnarray}
m_u^{}&=&1.27\,\textrm{MeV}\,,~m_c^{}=0.703\,\textrm{GeV}\,,~m_t^{}=168.7\,\textrm{GeV}~~~~\nonumber\\
m_d^{}&=&2.90\,\textrm{MeV}\,,~m_s^{}=40\,\textrm{MeV}\,,~~~\,~m_b^{}=2.98\,\textrm{GeV}\,,\nonumber\\
m_e^{}&=&0.486570161\,\textrm{MeV}\,,~~V_{ud}^{}=0.97419\,,\nonumber\\
m_\mu^{}&=&102.7181359\,\textrm{MeV}\,,~~\,V_{cs}^{}=0.97334\,,\nonumber\\
m_\tau^{}&=&1776.24\,\textrm{MeV}\,,~~~~~~~~~V_{tb}^{}=0.999133\,,\nonumber\\
&&\Delta m_{21}^2= 7.60 \times
10^{-5}_{}\,\textrm{eV}^2_{}\,,\nonumber\\
&&\Delta m_{31}^2 = 2.40\times 10^{-3}_{}\,\textrm{eV}^2_{}\,,
\end{eqnarray}
to derive
\begin{eqnarray}
\label{parameter2} \epsilon=1.60\times 10^{-9}_{}\,,
\end{eqnarray}
and then the normal hierarchy neutrino masses,
\begin{eqnarray}
m_{\nu_1^{}}^{}&=&1.69\times
10^{-3}_{}\,\textrm{eV}\,,~~m_{\nu_2^{}}^{}=8.88\times
10^{-3}_{}\,\textrm{eV}\,,\nonumber\\
m_{\nu_3^{}}^{}&=&4.90\times 10^{-2}_{}\,\textrm{eV}\,,~~
\sum_{i=1}^3 m_{\nu_i^{}}^{}=0.0596\,\textrm{eV}\,,
\end{eqnarray}
which are consistent with the cosmological limit \cite{dunkley2008}.

We further find the appearance of the relation (\ref{relation}) is
not accidental in a $SU(3)_c^{}\times SU(2)_L^{}\times
SU(2)_R^{}\times U(1)_{B-L}^{}$ left-right symmetric model
\cite{ps1974}. In our model, the scalar sector contains a real
singlet $\sigma(\textbf{1},\textbf{1},\textbf{1},0)$ and a
leptoquark singlet
$\delta(\textbf{3},\textbf{1},\textbf{1},-\frac{2}{3})$ as well as a
left-handed doublet $\phi_L^{}(\textbf{1},\textbf{2},\textbf{1},-1)$
and its right-handed partner
$\phi_R^{}(\textbf{1},\textbf{1},\textbf{2},-1)$. In the fermion
sector, besides the usual quark and lepton doublets, i.e.
$q_{L}^{}(\textbf{3},\textbf{2},\textbf{1},\frac{1}{3})$,
$q_{R}^{}(\textbf{3},\textbf{1},\textbf{2},\frac{1}{3})$,
$l_{L}^{}(\textbf{1},\textbf{2},\textbf{1},-1)$ and
$l_{R}^{}(\textbf{1},\textbf{1},\textbf{2},-1)$, we introduce four
types of singlets:
$D_{L,R}^{}(\textbf{3},\textbf{1},\textbf{1},-\frac{2}{3})$,
$U_{L,R}^{}(\textbf{3},\textbf{1},\textbf{1},\frac{4}{3})$,
$E_{L,R}^{}(\textbf{1},\textbf{1},\textbf{1},-2)$ and
$N_R^{}(\textbf{1},\textbf{1},\textbf{1},0)$. Here all fermions have
three generations and their family indices have been suppressed. The
left-right symmetry is assumed to be the charge-conjugation, under
which the fields transform as
\begin{eqnarray}
\label{lrsymmetry} &\sigma\leftrightarrow
-\sigma\,,~\phi_{L}^{}\leftrightarrow\phi_{R}^\ast\,,~\delta
\leftrightarrow \delta^\ast_{}\,,~q_{L}^{}\leftrightarrow
q_{R}^{c}\,,~D_{L}^{}\leftrightarrow
D_{R}^{c}\,,&\nonumber\\
&U_{L}^{}\leftrightarrow U_{R}^{c}\,,~l_{L}^{}\leftrightarrow
l_{R}^{c}\,,~E_{L}^{}\leftrightarrow
E_{R}^{c}\,,~N_{R}^{}\leftrightarrow N_{R}^{}\,.&
\end{eqnarray}
Here the charge-conjugation of the fermions is defined by
$q_L=P_L^{}q\leftrightarrow
q_R^c=(q_R^{})^c_{}=(P_R^{}q)^c_{}=P_L^{}q^c_{}$, etc. We also
impose a $U(1)^3_{}=U(1)_1^{}\times U(1)_2^{} \times U(1)_3^{}$
global symmetry, under which
$(q_L^{},q_R^c,D_L^c,D_R^{},U_L^c,U_R^{})$, $(l_L^{},l_R^c,N_R^{})$,
$(E_L^{},E_R^c)$ and $\delta$, respectively, carry the quantum
numbers $(1,0,1)$, $(0,1,-1)$, $(2,1,1)$ and $(1,1,0)$, while
$\sigma$ and $\phi_{L,R}^{}$ are trivial. Clearly, this global
symmetry is consistent with the left-right symmetry
(\ref{lrsymmetry}).

The full scalar potential is easy to read,
\begin{eqnarray}
\label{potential}
V&=&\frac{1}{2}\mu_\sigma^2\sigma^2_{}+\mu_\phi^2(|\phi_L^{}|^2_{}+|\phi_R^{}|^2_{})+\mu\sigma(|\phi_L^{}|^2_{}-|\phi_R^{}|^2_{})\nonumber\\
&&+\mu_\delta^2|\delta|^2_{}+\frac{1}{4}\lambda_\sigma^{}\sigma^4_{}
+\lambda_\phi^{}(|\phi_L^{}|^4_{}+|\phi_R^{}|^4_{})\nonumber\\
&&+\lambda'^{}_\phi|\phi_L^{}|^2_{}
|\phi_R^{}|^2_{} +\lambda_\delta^{}|\delta|^4_{}+\lambda_{\sigma\phi}^{}\sigma^2_{}(|\phi_L^{}|^2_{}+|\phi_R^{}|^2_{})\nonumber\\
&&+\lambda_{\sigma\delta}^{}\sigma^2_{}|\delta|^2_{}+\lambda_{\phi\delta}^{}(|\phi_L^{}|^2_{}+|\phi_R^{}|^2_{})|\delta|^2_{}\,.
\end{eqnarray}
Here the parity-odd singlet $\sigma$ is essential to realize the
spontaneous D-parity violation \cite{cmp1984}, which can guarantee
the breakdown of $SU(2)_R^{}\times U(1)_{B-L}^{}$ to $U(1)_{Y}^{}$.
As for the Yukawa interactions, only the following terms are
allowed,
\begin{eqnarray}
\label{yukawa} \hskip -0.5cm \mathcal{L}_Y^{}=
-y_D^{}(\bar{q}_L^{}\widetilde{\phi}_L^{}D_R^{}+\bar{q}_R^{c}\widetilde{\phi}_R^{\ast}D_L^{c})
-y_U^{}(\bar{q}_L^{}\phi_L^{}U_R^{}+\bar{q}_R^{c}\phi_R^{\ast}U_L^{c})&&\phantom{xxx}\nonumber\\
-y_N^{}(\bar{l}_L^{}\phi_L^{}N_R^{}+\bar{l}_R^c\phi_R^\ast
N_R^{})-h(\delta\bar{l}_L^{}i\tau_2^{}q_L^c+\delta^\ast_{}\bar{l}_R^c
i\tau_2^{}q_R^{})\,&&\nonumber\\
-f(\delta\overline{U}_L^{}E_L^c+\delta^\ast_{}\overline{U}_R^c
E_R^{})+\textrm{H.c.}\,.\quad\quad\quad\quad\quad\quad\quad\quad\quad\,&&\nonumber
\end{eqnarray}
\vspace{-1.52cm}
\begin{eqnarray}
\end{eqnarray}
We further introduce the mass terms of the fermion singlets by
softly breaking the $U(1)_{}^3$ global symmetry,
\begin{eqnarray}
\label{soft} \mathcal{L}_{soft}^{}&=&
-m_D^{}\overline{D}_L^{}D_R^{}-
m_U^{}\overline{U}_L^{}U_R^{}-m_E^{}\overline{E}_L^{}E_R^{}\nonumber\\
&&-\frac{1}{2} m_N^{}\overline{N}_R^c N_R^{}+\textrm{H.c.}\,,
\end{eqnarray}
where the mass matrices $m_D^{}$, $m_U^{}$, $m_E^{}$ and $m_N^{}$
are all symmetric, i.e.,
\begin{eqnarray}
m_D^{}=m_D^T\,, ~~m_U^{}=m_U^T\,, ~~m_E^{}=m_E^T\,,
~~m_N^{}=m_N^T\,.
\end{eqnarray}
Without loss of generality and for convenience we will choose the
base with the diagonal and real $m_D^{}$, $m_U^{}$, $m_E^{}$ and
$m_N^{}$.

\begin{figure}
\vspace{3.8cm} \epsfig{file=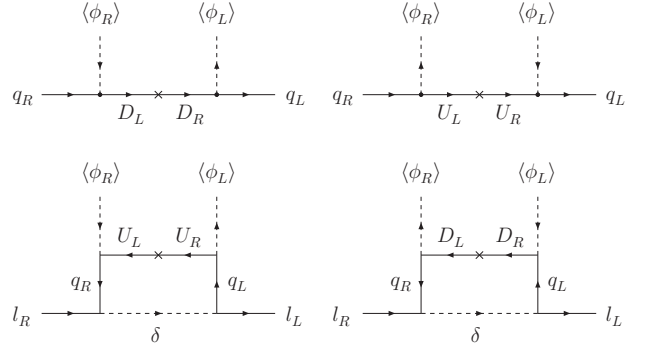, bbllx=5.3cm, bblly=6.0cm,
bburx=15.3cm, bbury=16cm, width=5.5cm, height=5.5cm, angle=0,
clip=0} \vspace{-4.5cm} \caption{\label{diracmass} Diagrams for
generating the Dirac masses of the down-type quarks (top left), the
up-type quarks (top right), the charged leptons (bottom left) and
the neutrinos (bottom right).}
\end{figure}

We now demonstrate that the quarks, the charged leptons, and the
neutrinos will obtain their Dirac masses,
\begin{eqnarray}
\hskip -.8cm \mathcal{L}\supset
-\tilde{m}_d^{}\bar{d}_L^{}d_R^{}-\tilde{m}_u^{}\bar{u}_L^{}u_R^{}
-\tilde{m}_e^{}\bar{e}_L^{}e_R^{}-\tilde{m}_\nu^{}\bar{\nu}_L^{}\nu_R^{}+\textrm{H.c.}\,,
\end{eqnarray}
after the symmetry breaking of $SU(2)_L^{}\times SU(2)_R^{}\times
U(1)_{B-L}^{}\stackrel{\langle\phi_R^{}\rangle}{\longrightarrow}
SU(2)_L^{}\times U(1)_{Y}^{}
\stackrel{\langle\phi_L^{}\rangle}{\longrightarrow} U(1)_{em}^{}$.
It is easy to check that the $f$-terms in the Yukawa couplings
(\ref{yukawa}) will contribute to the fermion masses only through
their radiative corrections to the masses of the leptoquark singlet
$\delta$ and the fermion singlets $U_{L,R}^{}$. Therefore, we will
not mention the $f$-terms in the following calculations and
discussions.

As shown in the tree-level diagrams of Fig. \ref{diracmass}, the
quark masses are induced by integrating out the colored fermion
singlets \cite{berezhiani1983},
\begin{eqnarray}
\hskip-.5cm\tilde{m}_d^{}=-y_D^{}\frac{v_L^{}v_R^{}}{m_D^{}}y_D^T=\tilde{m}_d^T\,,~
\tilde{m}_u^{}=-y_U^{}\frac{v_L^{}v_R^{}}{m_U^{}}y_U^T=\tilde{m}_u^T\,,
\end{eqnarray}
with
\begin{eqnarray}
v_R^{}=\langle\phi_R^{}\rangle\quad\textrm{and}\quad
v_L^{}=\langle\phi_L^{}\rangle\simeq 174\,\textrm{GeV}\,.
\end{eqnarray}
Note that the top quark mass is very close to $v_L^{}$. We thus need
$m_U^{}\sim v_R^{}\gg  v_L^{}$ to fulfill the perturbative and
seesaw \cite{minkowski1977} conditions. The choice is similar for
$m_D^{}$. The charged leptons and neutral neutrinos can not acquire
the Dirac masses at tree level. Instead, their Dirac masses are
generated at one-loop order, as shown in the box diagrams of Fig.
\ref{diracmass}. We calculate the loop-induced masses to be
\begin{eqnarray}
\label{emass}
\tilde{m}_e^{}=\frac{c_e^{}}{16\pi^2_{}}hy_U^\ast\frac{v_L^{}v_R^{}}{m_U^{}}y_U^\dagger
h^T_{}= -\frac{c_e^{}}{16\pi^2_{}}h\tilde{m}_u^\dagger h^T_{}=\tilde{m}_e^T\,,&&\\
\label{numass} \tilde{m}_\nu^{}=\frac{c_\nu^{}}{16\pi^2_{}}h
y_D^\ast\frac{v_L^{}v_R^{}}{m_D^{}}y_D^\dagger h^T_{}
=-\frac{c_\nu^{}}{16\pi^2_{}}h\tilde{m}_d^\dagger
h^T_{}=\tilde{m}_\nu^T\,.&&
\end{eqnarray}
Here the coefficients
\begin{eqnarray}
c_e^{}&=&\ln\frac{m_{U_i^{}}^2+m_\delta^2}{m_\delta^2}+\frac{m_{U_i^{}}^2}{m_\delta^2}\ln\frac{m_{U_i^{}}^2+m_\delta^2}{m_{U_i^{}}^2}\nonumber\\
&\simeq&1+\ln\frac{m_{U_i^{}}^2}{m_\delta^2}=\mathcal{O}(1-10)~~\textrm{for}~~m_\delta^2\ll m_{U_i^{}}^2\,,\\
c_\nu^{}&=&
\ln\frac{m_{D_i^{}}^2+m_\delta^2}{m_\delta^2}+\frac{m_{D_i^{}}^2}{m_\delta^2}\ln\frac{m_{D_i^{}}^2+m_\delta^2}{m_{D_i^{}}^2}\nonumber\\
&\simeq&1+\ln\frac{m_{D_i^{}}^2}{m_\delta^2}=\mathcal{O}(1-10)~~\textrm{for}~~m_\delta^2\ll
m_{D_i^{}}^2\,,
\end{eqnarray}
can be treated as constants since they are not very sensitive to the
running of $m_{U_i^{}}^2/m_\delta^2$ or $m_{D_i^{}}^2/m_\delta^2$.
The Dirac masses between the left- and right-handed neutrinos can be
determined by the charged fermion masses, i.e.
\begin{eqnarray}
\label{numass2} \hskip -0.5cm\tilde{m}_\nu^{}
=\frac{c_\nu^{}}{c_e^{}}U_e^{}\sqrt{\hat{m}_e^{}}\frac{1}{\sqrt{\hat{m}_u^{}}}V_{\textrm{CKM}}^\ast
\hat{m}_d^{}V_{\textrm{CKM}}^\dagger\frac{1}{\sqrt{\hat{m}_u^{}}}\sqrt{\hat{m}_e^{}}U_e^T\,.
\end{eqnarray}
Here $V_{\textrm{CKM}}^{}=V_u^{} V_d^\dagger$ is the CKM matrix
while $\hat{m}_d^{}$, $\hat{m}_u^{}$ and $\hat{m}_e^{}$ are the
diagonal mass matrices of the charged fermions,
\begin{eqnarray}
\hat{m}_d^{}&=&V_d^{}\tilde{m}_d^{}V_d^T=\textrm{diag}\{m_d^{},m_s^{},m_b^{}\}\,,\nonumber\\
\hat{m}_u^{}&=&V_u^{}\tilde{m}_u^{}V_u^T=\textrm{diag}\{m_u^{},m_c^{},m_t^{}\}\,,\nonumber\\
\hat{m}_e^{}&=&U_e^\dagger\tilde{m}_e^{}U_e^\ast=\textrm{diag}\{m_e^{},m_\mu^{},m_\tau^{}\}\,.
\end{eqnarray}
Note that we require the Yukawa couplings $h\lesssim 1$ for a
perturbative theory. With this constraint, the loop-induced charged
lepton masses can still arrive at the desired values. Actually, the
loop factor is expected to give the mass ratio between the tau
lepton and the top quark.

The completed neutrino mass terms are given by
\begin{eqnarray}
\mathcal{L}&\supset &
-\tilde{m}_\nu^{}\bar{\nu}_L^{}\nu_R^{}-y_N^{}v_L^{}\bar{\nu}_L^{}
N_R^{}-y_N^{} v_R^{}\bar{\nu}_R^c
N_R^{}\nonumber\\
&&-\frac{1}{2}m_N^{}\overline{N}_R^c
N_R^{}+\textrm{H.c.}\nonumber\\
&=&-\frac{1}{2}(\bar{\nu}_L^{},\bar{\nu}_R^c,
\overline{N}_R^c)\left(\begin{array}{c|cc}
0 ~~& ~~\tilde{m}_\nu^{} & y_N^{}v_L^{}\\
[2.5mm]\hline~&~&~\\
[-2.5mm]
\tilde{m}_\nu^{} & 0 & y_N^{} v_R^{}\\
[2.5mm] y_N^T v_L^{} & y_N^T v_R^{} & m_N^{}
\end{array}\right)\left(\begin{array}{c}
\nu_L^c \\
[4mm]\nu_R^{}  \\ [2.5mm]N_R^{}
\end{array}\right)\nonumber\\
&&+\textrm{H.c.}\,.
\end{eqnarray}
For $\tilde{m}_\nu^{}$ and $y_N^{}v_L^{}$ much smaller than $y_N^{}
v_R^{}$ and/or $m_N^{}$, we can make use of the seesaw
\cite{minkowski1977} formula,
\begin{eqnarray}
\mathcal{L}\supset
-\frac{1}{2}m_\nu^{}\bar{\nu}_L^{}\nu_L^c+\textrm{H.c.}\,,
\end{eqnarray}
where the mass matrix $m_\nu^{}$ contains two parts,
\begin{eqnarray}
\label{numass3}
m_\nu^{}=\tilde{m}_\nu^{}\frac{1}{y_N^T}m_N^{}\frac{1}{y_N^{}}\tilde{m}_\nu^{}\frac{1}{v_R^2}-2\tilde{m}_\nu^{}\frac{v_L^{}}{v_R^{}}\,.
\end{eqnarray}
The first term is the double \cite{mohapatra1986} or inverse
\cite{mv1986} seesaw whereas the second one is the linear
\cite{barr2003} seesaw. We further assume
\begin{eqnarray}
\label{assumption}
\frac{1}{y_N^T}m_N^{}\frac{1}{y_N^{}}\frac{1}{v_R^2}=2\frac{v_L^{}}{v_R^{}}\left(\frac{1}{\tilde{m}_\nu^{}}
+\frac{1}{\tilde{m}_\nu^{}}U_1^{}\tilde{m}_\nu^{}U_1^T\frac{1}{\tilde{m}_\nu^{}}\right)\,,
\end{eqnarray}
to parametrize the neutrino mass matrix (\ref{numass3}),
\begin{eqnarray}
\label{numass4} m_\nu^{}= 2\frac{v_L^{}}{v_R^{}}
U_1^{}\tilde{m}_\nu^{}U_1^T\,.
\end{eqnarray}
Consequently, we can perform
\begin{eqnarray}
\label{numass5} m_\nu^{}&=&
2\frac{v_L^{}}{v_R^{}}\frac{c_\nu^{}}{c_e^{}} (U_1^{}
U_e^{})\sqrt{\hat{m}_e^{}}\frac{1}{\sqrt{\hat{m}_u^{}}}V_{\textrm{CKM}}^\ast
\hat{m}_d^{}V_{\textrm{CKM}}^\dagger\nonumber\\
&&\times \frac{1}{\sqrt{\hat{m}_u^{}}}\sqrt{\hat{m}_e^{}}(U_e^T
U_1^T)\,.
\end{eqnarray}
With the charged fermion masses (\ref{fermionmass}) and the CKM
matrix (\ref{ckm}), it is easy to find
\begin{widetext}
\begin{eqnarray}
&&\sqrt{\hat{m}_e^{}}\frac{1}{\sqrt{\hat{m}_u^{}}}V_{\textrm{CKM}}^\ast
\hat{m}_d^{}V_{\textrm{CKM}}^\dagger\frac{1}{\sqrt{\hat{m}_u^{}}}\sqrt{\hat{m}_e^{}}\nonumber\\
[2mm] &\simeq&\left(\begin{array}{c|c}
~\displaystyle{\frac{m_e^{}}{m_u^{}}[m_d^{}(V_{ud}^{\ast})^2_{}+m_s^{}(V_{us}^{\ast})^2_{}+m_b^{}(V_{ub}^{\ast})^2_{}]}~~&
~~\begin{array}{cc}\quad\quad\displaystyle{\sqrt{\frac{m_e^{}m_\mu^{}}{m_u^{}m_c^{}}}m_s^{}V_{us}^\ast
V_{cs}^\ast}
& \quad\quad\quad\quad\displaystyle{\sqrt{\frac{m_e^{}m_\tau^{}}{m_u^{}m_t^{}}}m_b^{}V_{ub}^\ast V_{tb}^\ast}\end{array}~\\
[5mm]\hline~&~\\
\begin{array}{c}\displaystyle{\sqrt{\frac{m_e^{}m_\mu^{}}{m_u^{}m_c^{}}}m_s^{}V_{us}^\ast
V_{cs}^\ast} \\
[5mm]~\\
\displaystyle{\sqrt{\frac{m_e^{}m_\tau^{}}{m_u^{}m_t^{}}}m_b^{}V_{ub}^\ast
V_{tb}^\ast}\end{array} &
\begin{array}{c|c}~\displaystyle{\frac{m_\mu^{}}{m_c^{}}[m_s^{}(V_{cs}^{\ast})^2_{}+m_b^{}(V_{cb}^{\ast})^2_{}]}~~
&~~\displaystyle{\sqrt{\frac{m_\mu^{}m_\tau^{}}{m_c^{}m_t^{}}}m_b^{}V_{cb}^\ast
V_{tb}^\ast}~ \\
[5mm]\hline~&~\\
\displaystyle{\sqrt{\frac{m_\mu^{}m_\tau^{}}{m_c^{}m_t^{}}}m_b^{}V_{cb}^\ast
V_{tb}^\ast}&~~\displaystyle{\frac{m_b^{}}{m_t^{}}m_\tau^{}(V_{tb}^\ast)^2_{}}\end{array}
\end{array}\right)\nonumber\\
[5mm]
&\simeq&U_2^{}\textrm{diag}\left\{\frac{m_d^{}}{m_u^{}}m_e^{}(V_{ud}^\ast)^2_{}\,,\displaystyle{\frac{m_s^{}}{m_c^{}}m_\mu^{}(V_{cs}^\ast)^2_{}\,,
\frac{m_b^{}}{m_t^{}}m_\tau^{}(V_{tb}^\ast)^2_{}}\right\}U_2^T\,,
\end{eqnarray}
\end{widetext}
so that
\begin{eqnarray}
\label{numass6} m_\nu^{}&=&
2\frac{v_L^{}}{v_R^{}}\frac{c_\nu^{}}{c_e^{}} (U_1^{} U_e^{} U_2^{})
\textrm{diag}\left\{\displaystyle{\frac{m_d^{}}{m_u^{}}m_e^{}(V_{ud}^\ast)^2_{}\,,}\right.\nonumber\\
&&\left.\displaystyle{\frac{m_s^{}}{m_c^{}}m_\mu^{}(V_{cs}^\ast)^2_{}\,,\frac{m_b^{}}{m_t^{}}m_\tau^{}(V_{tb}^\ast)^2_{}}\right\}
(U_1^{}U_e^{} U_2^{})^T_{}\,.
\end{eqnarray}
The Pontecorvo-Maki-Nakagawa-Sakata (PMNS) \cite{mns1962} leptonic
mixing matrix is then given by
\begin{eqnarray}
U_{\textrm{PMNS}}^{}= U_e^\dagger U_\nu^{}= U_e^\dagger U_1^{}U_e^{}
U_2^{}\,.
\end{eqnarray}
For the given $U_{\textrm{PMNS}}^{}$ and $U_2^{}$, we can choose an
arbitrary unitary $U_e^{}$ to determine the Yukawa couplings
$y_N^{}$ and the Majorana masses $m_N^{}$ by inserting the unitary
$U_1^{}=U_e^{}U_\textrm{PMNS}^{}U_2^\dagger U_e^{}$ and the known
$\tilde{m}_\nu^{}$ [cf. (\ref{numass2})] to Eq. (\ref{assumption}).
The seesaw and perturbative conditions can be satisfied in a wide
parameter space of $y_N^{}$ and $m_N^{}$. Clearly, the neutrino mass
matrix (\ref{numass6}) can perfectly produce the mass spectrum
(\ref{spectrum}) for generating the relation (\ref{relation}).
Compared with Eqs. (\ref{spectrum}) and (\ref{parameter2}), we can
find
\begin{eqnarray}
\epsilon =2\frac{v_L^{}}{v_R^{}}\frac{c_\nu^{}}{c_e^{}}=1.60\times
10^{-9}_{}\,,
\end{eqnarray}
to determine the left-right symmetry breaking scale,
\begin{eqnarray}
v_R^{}= 2\frac{v_L^{}}{\epsilon}\frac{c_\nu^{}}{c_e^{}}=2.18\times
10^{11}_{}\,\textrm{GeV}~~
\textrm{for}~~\frac{c_\nu^{}}{c_e^{}}=1\,.
\end{eqnarray}

In summary we found the intriguing relation between the neutrino and
charged fermion masses, $|m_{\nu_3^{}}^2- m_{\nu_1^{}}^2| :
(m_{\nu_2^{}}^2- m_{\nu_1^{}}^2):: V_{tb}^4 m_\tau^2 m_b^2/m_t^2
 : V_{cs}^4 m_\mu^2 m_s^2/m_c^2$. We then proposed a left-right
symmetric model to naturally explain this phenomenon. In our model,
the normal hierarchy neutrino masses are fully determined by the
charged fermion masses for a given left-right symmetry breaking
scale. In turn, the left-right symmetry breaking scale is fixed by
the observed neutrino masses. The predicted neutrino spectrum is
possible to test in the future. Moreover, the leptoquark singlet
scalar in our model is flexible to be at an accessible scale so that
it can have some interesting implications on the present and future
experiments.

\vspace{1mm}

\textbf{Acknowledgement}: I thank Manfred Lindner for hospitality at
Max-Planck-Institut f\"{u}r Kernphysik. This work is supported by
the Alexander von Humboldt Foundation.

\end{document}